\documentstyle[manuscript,aps]{revtex}
\begin{document}

\title{Time dynamics in chaotic many-body systems:
can chaos destroy a quantum computer?}

\author{V.V. Flambaum \thanks{email address: flambaum@newt.phys.unsw.edu.au}}

\address{ School of Physics, University of New South Wales,
Sydney 2052, Australia}

%\twocolumn[
%\date{\today}
\maketitle
%\widetext

\tightenlines

%\vspace*{-1.0truecm}

\begin{abstract}
%\begin{center}
%\parbox{14cm}
  Highly excited many-particle states in quantum  systems (nuclei, atoms,
quantum dots, spin systems, quantum computers)
can be ``chaotic'' superpositions of mean-field
basis states (Slater determinants, products of spin or qubit states).
This is a result of the very high energy level  density of many-body
states which can be easily mixed by a  residual interaction between 
particles.
We consider the time dynamics of wave functions and increase of entropy in
such chaotic
systems.

 As an example we present the time evolution
 in a closed quantum computer.
 A time scale for the entropy $S(t)$ increase is $t_c
\sim \tau_0/(n \log_2{n})$, where   $\tau_0$ is the qubit ``lifetime'',
n is the number of qubits, $S(0)=0$ and $S(t_c)=1$.
At $t \ll t_c$ the entropy  is small: $ S  \sim n t^2 J^2 
\log_2(1/t^2 J^2)$, where $J$ is the 
inter-qubit interaction strength. At  $t > t_c$ the number of ``wrong'' states
increases exponentially as $2^{S(t)}$ . Therefore, $t_c$  may be
 interpreted as a maximal time
for operation of a quantum computer, since at $t > t_c$ one has to struggle
against the second law of thermodynamics. At $t \gg t_c$
 the system entropy approaches that for chaotic eigenstates.
 
%\end{center}
\end{abstract}


%] \narrowtext
\section{Introduction}
  Highly excited many-particle states in many-body systems
can be presented as ``chaotic'' superpositions of shell-model
basis states - see the recent calculations for complex atoms \cite{Ce},
multicharged ions \cite{ions}, nuclei \cite{nuclei} and
 spin systems \cite{Nobel,spins}. Indeed, the number of combinations
to distribute $n$ particles over $m$ orbitals is exponentially large
($m!/n!(m-n)!$  in a Fermi system). Therefore, the interval between
 the many-body levels $D$ is exponentially small and residual interaction
 between the particles mixes a huge number of the mean-field basis states
(Slater determinants) when forming eigenstates. The number of
principal basis components in an eigenstate can be estimated
as $N_p \sim \Gamma/D$ where $\Gamma$ is the spreading width
of a typical component that can be estimated using the Fermi Golden Rule.
In such chaotic eigenstates any external weak perturbation is
exponentially enhanced. The enhancement factor is $\sim \sqrt{N_p} \propto
1/\sqrt{D}$ - see e.g. \cite{enhancement} and references therein.
 This huge enhancement have
been observed in numerous experiments studying parity violation
effects in compound nuclei - see e.g. review \cite{W} and
 references therein.

  In a recent  work \cite{S} the consideration of many-body
chaos has been extended to quantum computers
\cite{feynman,shor1,shor2,steane1,steane2,zoller,nmr1,nmr2,vagner,kane,loss,cooper,lattice,monroe}.
 Any model of a quantum
 computer is somewhat similar to that of a spin system. In Ref.\cite{S}
the authors modelled a quantum computer by a random Hamiltonian,
\begin{equation}
\label{hamil}
H = \sum_{i} \epsilon_i \sigma_{i}^z + \sum_{i<j} J_{ij} 
\sigma_{i}^x \sigma_{j}^x,
\end{equation}
where the $\sigma_{i}$ are the Pauli matrices for the qubit $i$ and the second
sum runs
over nearest-neighbor qubit pairs.
The energy spacing between the two states of a qubit was represented 
by $\epsilon_i$ which was uniformly distributed in the interval 
$[0.5 \Delta_0, 1.5\Delta_0 ]$. 
Here $\epsilon_i$ can be viewed as the splitting
of nuclear spin levels in a local magnetic field, as  discussed 
in recent experimental
proposals \cite{vagner,kane}.  The different values of $\epsilon_i$ 
are needed to prepare a specific initial state by electromagnetic pulses
in nuclear magnetic resonance.  
In this case the couplings $J_{ij}$ will
represent the  interactions between the spins, which are needed
for multi-qubit operations in
 the quantum computer.  The total number of states in this system
is $N=2^n$, and the typical interval between the nearby energies of multiqubit
states is $\sim \Delta_0 n 2^{-n}$.

 A rough estimate
for the boundary of the chaos in the quantum computer eigenstates
is $ J_c \sim  \Delta_0/qn$,
where $ qn$ is the number of
interacting qubit pairs ($qn=2n$ in a 2D  square array of ``spins''
with only short-range interactions). This 
follows from a simple perturbation theory argument: the mixing is strong
when the perturbation is larger than the minimal energy interval
between the basis states which can be directly mixed by this
perturbation (see detailed discussion of the boundary of chaos
in many-body systems in Refs. \cite{Altshuler,FI97,FI99}).
 Numerical simulations in \cite{S} have shown
that the boundary of the chaos in the quantum computer eigenstates
is $ J_c \simeq 0.4 \Delta_0/n$ .
Above this point they observed a transition from  Poissonian to Wigner-Dyson
 statistics for the intervals between  the energy levels.
 For $J < J_c$ one eigenstate is formed
by one or few basis states built from the non-interacting qubits
(products of ``up'' and ``down'' states). For  $J > J_c$ a huge number
 of basis states are required.

 Because of the exponential laws
it is convenient to study the entropy S of the eigenstates
(in many-body systems the  entropy $S \simeq \ln{N_p}$,
 see e.g. Ref. \cite{FI97}). 
 In Ref. \cite{S} they observed a dramatic increase
of the eigenstate entropy in the transition
 from $J<J_c$ to $J> J_c$; in fact , they
defined $J_c$ as a point where $S=1$. In  Ref. \cite{S} this process
of chaotization of the eigenstates with the increase of $J$, or 
 number of qubits $n$, was termed as a melting of the quantum computer
and was assumed to lead to destruction of its operability.
The authors stress that this destruction of operability takes place
in an isolated  (closed) system without any external decoherence
process ( one could complement this picture by the $\sqrt{N_p}$
enhancement of any weak external perturbation acting on the quantum computer).

    This straightforward conclusion
may be misleading. ``Theoretically'', this picture is 
similar to that observed in nuclei and atoms. However, the ``experimental''
situation is very different. In nuclei and atoms  experiments
have resolved particular many-body energy levels. Therefore,
the description of the systems based on a consideration
of the eigenstates was an adequate one. In quantum computers the energy
interval between the  eigenstates is extremely small.
 The authors of Ref. \cite{S} estimated
that the average interval between the multi-qubit eigenstates
for 1000 qubits, the minimum number for which Shor's algorithm \cite{shor1}
 becomes useful \cite{steane2} is $D \sim 10^{-298} K$ (for a realistic 
$\Delta_0 \sim 1K$). Therefore, in the case of a quantum computer it is
 impossible to resolve multiqubit energy levels. Temperature, or the finite
time of the process $\tau$, gives an uncertainty in energy $\delta E \gg D$.
 In this case the picture with chaotic eigenstates is not an adequate
one and we should consider the time evolution of the quantum computer
wave function and entropy. Quantum chaos in the eigenstates allows us
to apply a statistical approach to this consideration.

\section{Time evolution of  the chaotic many-body state}
Exact (``compound'') eigenstates $\left| k\right\rangle \,$of the
Hamiltonian $H$ can be expressed in terms of simple shell-model basis states 
$\left| f\right\rangle $ in many-body systems or products of qubits
 in a computer:
\begin{equation}
\label{slat}\left| k\right\rangle =\sum\limits_f C_f^{(k)}\left|
f\right\rangle \,;\,\,\,\,\,\,\,\,\left| f\right\rangle
=a_{f_1}^{+}...a_{f_n}^{+}\left| 0\right\rangle. 
\end{equation}
These compound eigenstates $\left| k\right\rangle $ are formed by the
residual interaction $J$ ; $a_s^{+}$ are creation or spin-raising operators
(if the ground state $\left| 0\right\rangle$ corresponds to spins down).
Consider now the time evolution of the system.
Assume that initially ($t=0$) the system is in a basis state
  $\left| i\right\rangle $ (quantum computer in
a state with certain  spins ``up'')
which can be presented as a sum over exact eigenstates:
\begin{equation}
\label{in}\left| i\right\rangle =\sum\limits_kC_i^{(k)}\left|
k\right\rangle. 
\end{equation}
Then the time-dependent wave function is equal to
\begin{equation}
\label{psit}\Psi (t) =\sum\limits_{k,f}C_i^{(k)}C_f^{(k)}\left|
f\right\rangle \exp(-i E^{(k)}t).
\end{equation}
 The sum is taken over the eigenstates $k$ and basis states $f$
; we  put $\hbar=1$.
 The probability $W_i=|A_i|^2 =|\left\langle i|\Psi(t)\right\rangle|^2$ to find
the initial state in this wave function is determined by the amplitude
\begin{equation}
\label{ampli}
A_i= \left\langle i|\exp(-iHt)|i\right\rangle=
\sum\limits_k|C_i^{(k)}|^2\exp(-i E^{(k)}t) \simeq
\int dE P_i(E)\exp(-i Et).
\end{equation}
Here we replaced the summation over a very large number of the eigenstates
by the integration over their energies $E \equiv E^{(k)}$ and introduced
the ``strength function'' $P_i(E)$ which is also known in the literature as
the ``local spectral density of states'',
\begin{equation}
\label{strength}P_i(E)\equiv \overline{|C_i^{(k)}|^2}\rho (E),
\end{equation}
where $\rho (E)$ is the density of the eigenstates. In chaotic systems
the strength function is given by a Breit-Wigner- type formula \cite{BM,FI99}:
\begin{equation}
\label{FfBW} P_i(E) = \frac{1}{2\pi}\frac{\Gamma_i(E)}{(E_i+\delta_i-E)^2 +
(\Gamma_i(E)/2)^2} ,
\end{equation}
\begin{equation}
\label{GammaH}\Gamma _i(E)\simeq 2\pi \overline{\left| H_{if}\right| ^2}\rho
_f(E) \sim  J^2 qn/\Delta_0 .
\end{equation}
Here $\delta_i$ is the correction to the unperturbed energy level $E_i$
due to the residual interaction $J$,  $\rho_f(E) \sim qn/\Delta_0$
 is the density of the ``final'' basis states directly connected
by the interaction matrix element  $H_{if}$ with the initial state
$\left| i\right\rangle$.
 We see from the equations above that the time
dynamics is determined by the structure of the eigenstates.  

  It is easy to find $W_i(t)$ for a small time $t$.  Let us separate 
the energy of the initial state $E_i\equiv H_{ii}$ in the  exponent
and make a  second order expansion in $H-E_i$ or $E-E_i$ in eq.(\ref{ampli}).
The result is:
\begin{equation}
\label{At2}
A_i= \exp(-i E_i t)(1- (\Delta E)^2 t^2/2) ,
\end{equation}
\begin{equation}
\label{Wt2}
W_i(t)= (1- (\Delta E)^2 t^2) ,
\end{equation}
\begin{equation}
\label{DeltaE}
(\Delta E)^2= \sum\limits_{f \neq i} H_{if}^2 = \sum\limits_{i<j}J_{ij}^2
= q n J_r^2 .
\end{equation}
Here $(\Delta E)^2$ is the second moment of the strength function,
$J_r$ is the r.m.s. value of the interaction strength,
$J_r^2\equiv\overline{J_{ij}^2}$.
The first moment is equal to $E_i=H_{ii}$, see e.g. \cite{FI97,FI99}
where one can also find the calculations of $(\Delta E)^2$
and spreading width $\Gamma(E)$ for many-body systems.

 Note that in the special case
of a very strong residual interaction $J \gg \Delta_0$ this short-time
dependence can be extended to a longer time 
using the exact solution for the case of $\Delta_0=0$
(in this case it is easy to calculate $\exp(-iHt)$):
\begin{equation}
\label{Alarge}
A_i=\prod\limits_{i<j} \cos{J_{ij}t}
 \simeq \prod\limits_{i<j}(1-(J_{ij}t)^2/2) \simeq \exp(-(\Delta E)^2 t^2/2), 
\end{equation}
\begin{equation}
\label{Wlarge}
W_i(t)\simeq\exp(- (\Delta E)^2 t^2).
\end{equation}
The strength function and density of states in this limit
are also described by  Gaussian functions
 with  variance $\sigma^2=(\Delta E)^2$:
\begin{equation}
\label{PGauss} P_i(E) = \frac{1}{\sqrt{2\pi \sigma^2}}
 \exp(-\frac{E^2}{2\sigma^2}),
\end{equation}
\begin{equation}
\label{rhoGauss} \rho(E) = \frac{2^n}{\sqrt{2\pi \sigma^2}}
 \exp(-\frac{E^2}{2\sigma^2}).
\end{equation}
 The density of states remains Gaussian for $\Delta_0 \neq 0$,  with $\sigma^2=
n \overline{\epsilon^2} + (\Delta E)^2$ if there is no gap
in the single-qubit spectra (in Ref. \cite{S} the ``up'' and ``down'
spectra were separated by a gap equal to $\Delta_0$).
In general  the unperturbed density of
states ($J=0$) can be presented as a sum of the Gaussian functions
(one should separate classes of states with a certain number of spins ``up'').
The interaction $J$ in the Hamiltonian (\ref{hamil}) mixes these classes
and makes the density closer to the single Gaussian function.

  The limit at large time in chaotic case  can be obtained by 
calculation of the integral in eq. (\ref{ampli}) in the 
 complex $E$ plain. We should close
 the contour of integration in the bottom part of the complex plane
($Im(E)<0$) to provide a vanishing contribution at infinity.
 The limit at large time $t$ is given by the
pole of the strength function  (\ref{FfBW})
closest to the real $E$ axis . If $\Gamma$ and $\delta_i$ do not depend
on $E$ the integration gives the usual exponential decay $W_i=\exp(-\Gamma t)$
 \cite{BM}.
However, the dependence of the spreading width on
 energy $E$ is necessary to provide the finite second moment $(\Delta E)^2$ 
of the strength function (note that in many-body systems the dependence
$\Gamma (E)$ can be approximated
by a Gaussian function, since the density of final states
$\rho_f(E_f)$ in eq. (\ref{GammaH}) is usually close to  
Gaussian \cite{FI99}).
If $\Gamma <  \Delta E$ the closest pole is given by $\tilde\Gamma=
- 2 Im(E_p)$, where $E_p$ is a solution of the equation
 $E_p=E_i +\delta_i(E_p) -
i \Gamma(E_p)/2$ with a minimal imaginary part.
If $\Gamma \ll \Delta E$ we have  $\tilde\Gamma=\Gamma$.
 As a result we obtain an exponential dependence for large t:
\begin{equation}
\label{Wtinf}
W_i(t) \sim \exp(- \tilde\Gamma t).
\end{equation}
  It is useful to have a simple extrapolation formula (valid for $\Gamma <
\Delta E$) between the cases
of small time eq. (\ref{Wt2}) and large time  eq. (\ref{Wtinf}):
\begin{equation}
\label{Wint}
W_i(t)=\exp\left(\frac{\Gamma^2}{2 (\Delta E)^2}-
 \sqrt{\frac{\Gamma^4}{4 (\Delta E)^4} +\Gamma^2t^2} \right) . 
\end{equation}

   Now we can estimate the probabilities of the other components
$W_f$. For small time or small interaction $J$, other components
can be populated due to  direct transitions from the initial state only: 
\begin{eqnarray}
\label{Wf}
W_f= |\left\langle f|\exp(-iHt)|i\right\rangle|^2 \simeq
|H_{if}|^2\left|\int\limits_0^t |A_i(t)| \exp(i\omega_{if}t)dt\right|^2 
\nonumber \\
\simeq\frac{|H_{if}|^2}{\omega_{if}^2+\Gamma^2/4}\left|\exp{(i\omega_{if}-\Gamma/2)t}
-1\right|^2 .
\end{eqnarray}
Here $\omega_{if}=E_f-E_i$. We stress again that this
approximate  equation does not contain transitions between the small components.
For example, it does not contain the width of the state $f$;
the width $\Gamma$ stands only to indicate some increase of the denominator
and to clarify the ``small time'' condition that should include 
small $\omega_{if}t$,  $\Gamma t$ or $\Delta E t$.
For small time $W_f=|H_{if}|^2 t^2$. Here $H_{if}$ is equal to one
of the $J_{ij}$ that produces a change of the state of a pair of 
``spins'' (qubits), transferring initial state $i$ to another state $f$.
 The result at larger times is different for perturbative and chaotic 
regimes. In the perturbative regime,
 $J \ll \Delta_0/qn$ 
 eq. (\ref{Wf}) is the final one. In the chaotic regime we can find
the asymptotic expression for  large times. The projection
of  $\Psi (t)$ in eq. (\ref{psit}) to the component $f$ gives
\begin{equation}
\label{sfluct}
W_f(t) =W_f^s + W_f^{fluct}(t),
\end{equation}
\begin{equation}
\label{ws}
W_f^s =\sum\limits_k|C_i^{(k)}|^2|C_f^{(k)}|^2 \simeq
\int \frac{dE}{\rho(E)} P_i(E) P_f(E) \simeq
\frac{1}{2\pi\rho}\frac{\Gamma_t}{(E_i-E_f)^2 +
(\Gamma_t/2)^2} .
\end{equation}
Here $\Gamma_t \simeq \Gamma_i+\Gamma_f  \simeq 2\Gamma$.
\begin{equation}
\label{fluc}
 W_f^{fluct}(t)=\sum\limits_{k,p;k\neq p}C_i^{(k)}C_f^{(k)}C_i^{(p)}C_f^{(p)}
 \exp(i(E^{(k)}-E^{(p)})t).
\end{equation}
 At large 
time $t$, the different terms in $ W_f^{fluct}(t)$ rapidly oscillate
and we can put  $ \overline{W_f^{fluct}(t)}=0 $. Thus, asymptotically
the distribution of the components in the time-dependent
wave function is close to
that in the chaotic eigenstates (see eqs (\ref{strength},\ref{FfBW}))
 with a doubled spreading width.  

\section{Entropy increase}

It is convenient to define the entropy of a many-body state
 as a sum over the basis components ( a comparison with other definitions
can be found, e.g. in Ref. \cite{FI97}):  
\begin{equation}
\label{entropy}
S = -\sum\limits_s W_s \log_2 W_s=
- W_i \log_2 W_i -\sum\limits_{f\neq i} W_f \log_2 W_f .
\end{equation}
 Initially, we have only one component, $W_i=1$ and the entropy is
 equal to zero. It is easy to obtain a small-time estimate for the entropy  
using eqs. (\ref{Wt2}, \ref{DeltaE}, \ref{Wf}):
\begin{equation}
\label{Ssmall}
S \simeq (\Delta E)^2 t^2  \log_2(qn/(\Delta E)^2 t^2) = qn J_r^2 t^2
 \log_2(1/J_r^2 t^2) .
\end{equation}
We see that the initial increase of the entropy is relatively small
($\sim t^2$), however, it is proportional to the number of qubits
$n$.

 The criterion of a quantum computer ``melting'' used in 
Ref. \cite{S} is the entropy $S=1$. We can extend the small-
time consideration to include this point.
For small time we  have some decrease of the initial component 
and population of the components  directly coupled to the 
initial one. The number of such small components is equal to the number
of  interacting pairs (qn) in the  Hamiltonian eq.(\ref{hamil}),
since each pair can change its state due to interaction and
this leads to a different many-body state.
 Using the normalization condition
$\sum_s W_s=1$ we obtain an estimate $\overline{W_f}=(1-W_i)/n_f$
where $n_f$ is the ``principal'' number of the final components.
Initially $n_f=qn$.
This gives us the following approximate expression for the entropy:
\begin{eqnarray}
\label{entropyi}
S = - W_i \log_2 W_i -  \log_2((1-W_i)/n_f) 
\sum\limits_{f\neq i} W_f \nonumber \\
=- W_i \log_2 W_i - (1-W_i) \log_2((1-W_i)/n_f) \simeq (1-W_i) 
 \log_2(n_f) .
\end{eqnarray}
The last approximate expression is an estimate with logarithmic
accuracy, assuming  $\log_2(n_f)$ is large.

The condition $S=1$ combined with eq.(\ref{Wint}) for $W_i(t)$ and
eq.(\ref{entropyi}) for the entropy $S(t)$ gives
\begin{equation}
\label{melt}
W_i(t)=\exp\left(\frac{\Gamma^2}{2 (\Delta E)^2}-
 \sqrt{\frac{\Gamma^4}{4 (\Delta E)^4} +\Gamma^2t^2}\right)= 1- 1/ \log_2(n_f).
\end{equation}
This means that the ``melting'' happens when the probability
to be in the initial state $W_i$ is still close to 1
(since $\log_2(n_f)$ is large).
The loss of operability of the quantum computer
is due to the admixture of a large number of
the small components (``wrong'' basis states).

 We should note that , strictly speaking, the argument of the $\log_2$ may
differ from $n_f=qn$, since the point $t_c$ can be outside the small
time approximation. However, the estimate in eq. (\ref{melt})
with $\log_2n_f \simeq \log_2n $ is valid with  logarithmic accuracy
(for example, a more accurate estimate in the case of $\Gamma \ll \Delta E$
is $n_f \simeq qn \Gamma/\Delta_0$; this follows from eq.(\ref{Wf})).

 Equation (\ref{melt}) allows us to obtain a simple
estimate for the maximal operational time $t_c$:
\begin{equation}
\label{tcc}
t_c\simeq\frac{\hbar}{\Gamma  \log_2(n)}\sqrt{1+\frac{\Gamma^2 \log_2n}
{(\Delta E)^2}}.
\end{equation}
In the case of $\Gamma \ll \Delta E$  we have
\begin{equation}
\label{tc}
t_c\simeq\frac{\hbar}{\Gamma  \log_2(n)}= \frac{\tau_0}{n \log_2(n)}
\end{equation}
Here $\tau_0= \hbar/\Gamma_0$ is the ``lifetime''
related to a single qubit, $\Gamma_0 = \Gamma/n$;
recall that $\Gamma$ is proportional  to the number of qubits n.
More accurate result can be obtained numerically
using expressions for $W_i$ and $W_f$ presented above.

   At this point we can say something about the effects of the environment.
They also lead to  ``depolarization'' of  a qubit,
which means nonzero probability of the opposite spin state.
If this probability is small we can speak about the probabilities
of the population of $n$ many-qubit basis states. Each
admixed basis state in this case has one of the qubit states different
from the initial state. 
To account for this effect one may use a real (experimental) qubit
lifetime $\tau_0$ in the estimate (\ref{tc}).

  For   $t > t_c$ the higher orders in $H_{if}^2 t^2$ expansion
become important and the number of the small components increases
exponentially: each state generates $qn$ new states. This
corresponds to an approximately linear increase in the entropy.
At $t \gg t_c$ we can use the asymptotic form (\ref{ws}) of the 
component  distribution. It is two times broader
($\Gamma_t= 2 \Gamma$) than the basis component distribution
of chaotic stationary states. This means that the
asymptotic  number of the principal components 
is equal to $N_p(t)=2 N_p^{(k)}$, where $N_p^{(k)} \sim \Gamma/D$ 
is the number of  principal components in a chaotic eigenstate.
It is easy to calculate the entropy in this case.
From the normalization condition $\sum_s W_s=1$, it follows
that $\overline{W_s}=1/N_p$. Then     
\begin{equation}
\label{entropyinf}
S = -\sum\limits_s W_s \log_2 W_s \simeq  \log_2 N_p\sum\limits_s W_s=
\log_2 N_p.
\end{equation}
Thus, the asymptotic value of the entropy
is $S(t \gg t_c)=\log_2(2 N_p^{(k)})= S^{(k)} +1$,
where $S^{(k)}= \log_2 N_p^{(k)}$ is the entropy of a chaotic eigenstate.
Note, that it is smaller than the maximal possible 
entropy $S_{max}=\log_2 2^n= n$. This is due to
localization of the wave function within the energy shell
 centered at the energy  
of the initial state $E_i$ with the width $2\Gamma$.  

\section{Conclusion}
  The time dependence  of the closed quantum computer wave function
is  different in the non-chaotic and chaotic regime.
In the non-chaotic case $ J \ll \Delta_0/n$, the number of principal
components $N_p \simeq 1$ and the wave function
remains localized near the initial state (as it was pointed
out in \cite{S} the  energy level density  of the 
many-qubit states can be exponentially
high even in this case). An increase in the number of qubits $n$
leads to a transition to a chaotic regime where $ J > \Delta_0/n$.
In this case one can operate the quantum computer within a limited
time $t < t_c=\tau_0/n \log_2 n$,
where $\tau_0$ is the ``lifetime''
of one qubit. For $t > t_c$ it is hardly possible to operate
the quantum computer, since in this case one faces a hopeless struggle
 against the
second law of  thermodynamics: increase of the entropy  $S(t)$
and very fast exponential increase of the number 
of  ``wrong'' states $N_p(t)= \exp S(t)$. The asymptotic value of
 the entropy is then close to that for chaotic eigenstates. 

  A similar picture for the entropy increase is expected in other
many-body systems. For example, one can consider a decay of 
a single-electron wave function in a many-electron quantum dot.
In this case  $t_c \sim \tau /\log_2 n_f$ where $n_f$ is the
effective number
of final states that contribute to the decay width $\Gamma=\hbar/\tau$.
One may also speculate about the ``entropy'' increase
for decay of a single-particle wave function in chaotic
quantum billiard or disordered system using expansion
of this wave function in the plane wave basis or the orbital
angular momentum basis.

This work was supported by the Australian Research Council. The
author is grateful to M.Yu. Kuchiev for
valuable discussion and to A.S. Dzurak for careful reading of the manuscript.

%\end{multicols}
\end{document}